\documentclass[sigconf]{acmart}
\usepackage{lipsum}

\AtBeginDocument{%
  \providecommand\BibTeX{{%
    \normalfont B\kern-0.5em{\scshape i\kern-0.25em b}\kern-0.8em\TeX}}}

\copyrightyear{2023} 
\acmYear{2023} 
\setcopyright{acmlicensed}\acmConference[WWW '23 Companion]{Companion Proceedings of the ACM Web Conference 2023}{April 30-May 4, 2023}{Austin, TX, USA}
\acmBooktitle{Companion Proceedings of the ACM Web Conference 2023 (WWW '23 Companion), April 30-May 4, 2023, Austin, TX, USA}
\acmPrice{15.00}
\acmDOI{10.1145/3543873.3587650}
\acmISBN{978-1-4503-9419-2/23/04}

\usepackage{subcaption}
\usepackage{mwe}

\begin{document}
\newcommand{\Secref}[1]{\S\ref{#1}}
\newcommand{\Figref}[1]{Fig.~\ref{#1}}

\title[Analyzing Activity and Suspension Patterns of Twitter Bots Attacking Turkish Twitter Trends]{Analyzing Activity and Suspension Patterns of Twitter Bots Attacking Turkish Twitter Trends by a Longitudinal Dataset}



\author{Tuğrulcan Elmas}
\affiliation{
  \institution{Indiana University}
  \city{Bloomington}
  \country{U.S.A.}}
\email{telmas@iu.edu}



\renewcommand{\shortauthors}{Tuğrulcan Elmas}






\begin{abstract}

Twitter bots amplify target content in a coordinated manner to make them appear popular, which is an \textit{astroturfing attack}. Such attacks promote certain keywords to push them to Twitter trends to make them visible to a broader audience. Past work on such fake trends revealed a new astroturfing attack named \textit{ephemeral astroturfing} that employs a very unique bot behavior in which bots post and delete generated tweets in a coordinated manner. As such, it is easy to mass-annotate such bots reliably, making them a convenient source of ground truth for bot research. In this paper, we detect and disclose over 212,000 such bots targeting Turkish trends, which we name astrobots. We also analyze their activity and suspension patterns. We found that Twitter purged those bots en-masse 6 times since June 2018. However, the adversaries reacted quickly and deployed new bots that were created years ago. We also found that many such bots do not post tweets apart from promoting fake trends, which makes it challenging for bot detection methods to detect them. Our work provides insights into platforms' content moderation practices and bot detection research. The dataset is publicly available at \url{https://github.com/tugrulz/EphemeralAstroturfing}.

\end{abstract}

\maketitle

\section{Introduction}

Bots on social media engage in malicious activities such as amplifying harmful narratives, spreading disinformation, or astroturfing. Platforms proactively or retrospectively remove such accounts due to violating their terms of service through inauthentic activity. However, platforms' efforts may fall short of countering such activity if they do not have enough moderation power or incentive to enforce their rules. Additionally, such bots may be exploiting vulnerabilities in the platforms. In such a case, removing bots does not prevent the threat, as adversaries can always employ new accounts to exploit the vulnerability.

Previous work on Twitter bots showed that Twitter is vulnerable to bot attacks. Recent work by Pfeffer et al.~\cite{pfeffer2023just} estimated that 20\% of the active Twitter users may observe bot behavior in September 2021. The bots do not only simulate genuine user activities in an automated manner but also attack platforms by exploiting their vulnerabilities. In 2021, we introduced such an attack named "Ephemeral Astroturfing" in which Twitter bots mention a slogan in a coordinated manner to make it appear popular that Twitter will list it on the trends and make it visible to all Twitter users~\cite{elmas2021ephemeral}. We define the attack as "ephemeral" because the accounts delete their tweets that originate the trend immediately after posting, even before the slogan reaches the trends list. The trending algorithm does not check for such deletion activity and continues to list the trend, which is a vulnerability. Due to this vulnerability, the source of the trends goes unnoticed by researchers and they cannot detect such bots as it is very challenging to collect the data associated with attacks. In fact, we found that a fake trend containing hate \#SuriyelilerDefolsun (Syrians Get Out) poisoned at least five social media studies because the researchers could not collect the tweets by bots who created the trend and attributed the trend to the genuine users~\cite{elmas2021ephemeral}.

Although we disclosed the attacks to Twitter twice, they are still ongoing as of 2023 and they employ thousands of bots. We observe that they are currently employed in Turkey, Brazil and the U.S. Among those, Turkey is the most targeted country. By extending our previous work, we detected 29,000 fake trends between 2015-2022 created by 212,000 bots. The fake trends created by these attacks can make up at least 52\% of the trends in a day. In doing so, we provide a longitudinal and extensive study of \emph{entire bot-nets}, revealing the arms race between adversaries running the bots and Twitter content moderation, which is a novel contribution to social media bot research. Precisely, we found that Twitter purged the trend bots 7 times, where they mass removed or locked at least 2000 bots in a month. However, we found that the adversaries reacted quickly by replacing those accounts with new bots. We also characterized the trend bots to provide insights for bot detection methods. We find that a significant portion of bots is silent; they do not post tweets other than the ones promoting fake trends. We disclose the bots along with the times they attack, which is novel. Our dataset and analysis will guide future research on the analysis of bot nets and bot detection on social media in general.  

Our contributions are as follows:

\begin{enumerate}
    \item We detect and report the prevalence of fake trends created by ephemeral astroturfing in Turkey between 2020-2022. (\Secref{sec:faketrendprevalence})
    \item We perform a longitudinal analysis of bot activity and suspensions, shedding light on the arms race between Twitter's content moderation and the adversaries. (\Secref{sec:astrobotcreationsuspension}). 
    \item We characterize the bots by their time and volume of activity. We find that bots attack multiple times and some are only used for attacks and stayed inactive otherwise (\Secref{sec:botcharacteristics}).
    \item We disclose a dataset of 212,000 bots and their activity. To the best of our knowledge, this is the first bot dataset that reports \emph{when} the accounts were bots (\Secref{sec:dataset}). 
    
\end{enumerate}

\section{Background \& Related Work}

We motivate our work by providing a brief survey on related work and highlighting our contributions. 

\subsection{Platform Vulnerabilities}

Social media platforms may have vulnerabilities that are exploited by adversaries to manipulate social media and engage in malicious activities such as spreading disinformation~\cite{mirza2023tactics}. The platforms may not always prevent the threat. In such a case, adversaries may continue to exploit them and researchers may track their behavior to understand their objectives and impact. For instance, Elmas et al.~\cite{elmas2020misleading} detected and analyzed accounts that are compromised or purchased and then secretly repurposed to use for another purpose while maintaining their followers. Torres et al.~\cite{torres2022manufacture} detected and analyzed follow trains in which users secretly grow their followers by promoting each other in public tweets and then later deleting those tweets. In this study, we focus on such an attack named ephemeral astroturfing in which bots create fake Twitter trends that consistently exploit a vulnerability in the platform's trending algorithms~\cite{elmas2021ephemeral}. As this attack is still not prevented by Twitter, we are able to mass detect the bots employed, analyze Twitter's content moderation practices through their aggregated activity, and be able to use them as ground truth for bot research. The attacks threaten the integrity of Twitter trends, civic integrity, and the validity of social media studies~\cite{elmas2023impact}.

\subsection{Twitter Bot Detection}

As social media platforms may not timely detect and remove potentially harmful accounts, researchers proposed methods to detect and analyze such accounts independently. Such accounts include users with political agenda~\cite{darwish2017seminar}, trolls~\cite{saeed2022trollmagnifier}, and social media bots, which we focus extensively on in this paper. Past researched focus on detecting bots using their profile features~\cite{lee2011seven,miller2014twitter, golbeck2019benford, yang2020scalable}, content features~\cite{wei2019twitter}, graph features~\cite{ali2019detect} and temporal activity~\cite{chavoshi2016debot,minnich2017botwalk}, or a combination of them~\cite{sayyadiharikandeh2020detection, kudugunta2018deep,efthimion2018supervised}. To provide ground truth for these detection methods, researchers use human annotation~\cite{yang2020scalable}, honeypots~\cite{lee2011seven} or purchase the bots' services from the adversaries controlling them~\cite{cresci2015fame,golbeck2019benford}. In this study, we provide ground truth by detecting a very unique and anomalous bot behavior that exploits a vulnerability of Twitter. As this behavior employs coordinated tweets that are deleted immediately, bot detection methods cannot detect this behavior unless they know those bots apriori and collect their tweets and deletions in real time. Thus, we employ our own detection method to detect them. 

\subsection{Fake Twitter Trends}

Social media platforms may designate popular content and further amplify them~\cite{elmas2023measuring}. Twitter trends are one such amplification mechanism that shows the popular topics on the platform, making them visible to a broad audience. Recent studies found that they have the capability to influence people's perceived social agenda and shape their personal agenda~\cite{zhang2023trendingnow, schlessinger2021effects}. As such, they attract adversaries who employ bots to push their slogans to Twitter trends. So far, systematic trend manipulation is studied in Turkey~\cite{elmas2021ephemeral}, Pakistan~\cite{kausar2022push}, and India~\cite{jakesch2021trend}. We specifically focus on Turkey because there is hard evidence that adversaries use bots to create fake trends.

\section{Preliminaries}

\begin{figure}[!htb]

    \begin{subfigure}{.51\linewidth}
        \centering
        \includegraphics[width=\linewidth]{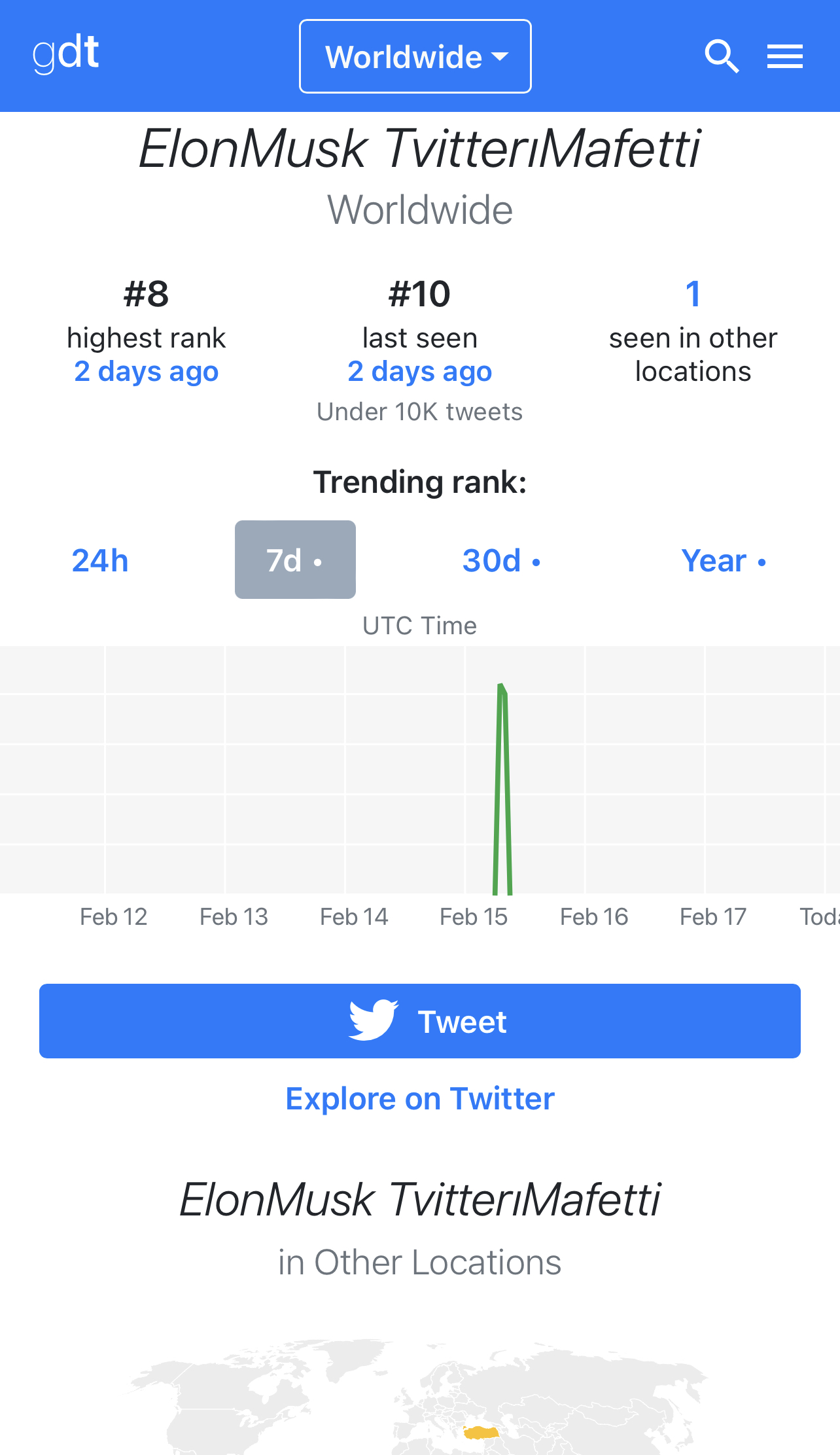}
    \end{subfigure}
    \begin{subfigure}{.48\linewidth}
        \centering
        \includegraphics[width=\linewidth]{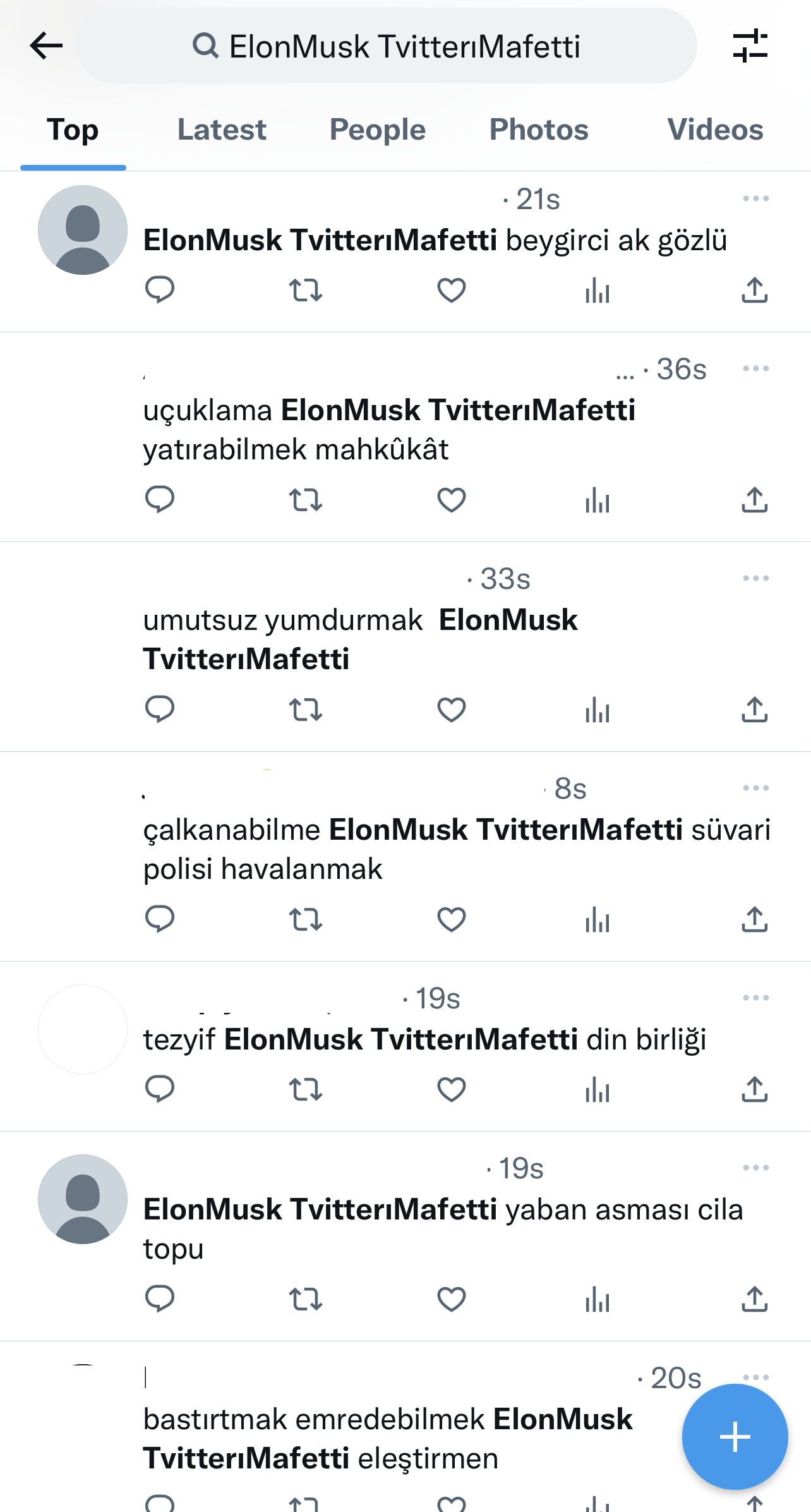}
    \end{subfigure}
    \caption{Figure (left) shows getdaytrends.com, a website dedicated to publishing Twitter trends worldwide, reports that ElonMusk TvitterıMafetti (ElonMusk RuinedTwitter) was trending worldwide on February 15 due to a single anomalous spike of activity. The only other location it was trending was Turkey. Figure (right) shows the bots pushing the trend by tweets randomly generated by picking up words from a Turkish lexicon. The tweets are already deleted.}
\label{fig:elonmusk}
\end{figure}

Social media platforms like Twitter employ popularity sensors such as the trending algorithm to assess the popularity or trendiness of social media content. Adversaries may employ fake accounts to interact with some content to give the impression that the content is popular, a deceptive strategy that is similar to political astroturfing~\cite{keller2020political}. Thus, we name \emph{astroturfing attack} as employing multiple and inauthentic accounts to perform \emph{actions} in a coordinated manner to deceive the popularity sensors. If the actions are later removed, but the impact stays intact, the attack is \textit{ephemeral}. 

Ephemeral astroturfing attacks may target and promote users, tweets, or some specific content. In this work, we focus on hashtags and n-grams that appear on Twitter trends in Turkey due to high volume of attacks in this country We name the trends that are promoted by those attacks as \emph{fake trend} and refer to them as such. 

Ephemeral astroturfing may be conducted using any kind of content that successfully simulate user-generated content to pass the spam filters of Twitter or any other bot detection mechanisms. Since such content will be later deleted, their function is mainly, if not only, to promote others. As such, we observe that the adversaries attacking Turkish trends on Twitter employ two very unique tweeting patterns that allow us to detect and mass annotate fake trends and the bots creating them. First, the tweets' content is generated by picking up random words from a Turkish lexicon and appending the target keyword (e.g., "critical to be able to boil lightning rod \#FakeTrend"). We observe that the tweets do not repeat themselves, which may increase the success of the attacks. We name these tweets as \emph{lexicon tweets}. They do not appear to have any function apart from mentioning the target keyword as they are meaningless. Second, they are posted in bulk by a set of Twitter accounts within a small time period (usually, within a minute) and then immediately deleted even before the target keyword reaches the trends list, making the attack ephemeral. The same set of Twitter accounts is used for multiple attacks.

Since lexicon tweets pushing fake trends are meaningless to humans, require a high level of coordination, and are very prevalent, they strongly indicate automated and inauthentic behavior, implying the authors of such tweets are bots and are centrally managed. We provide two instances of hard evidence to show that this is really the case. First, Twitter declared that they removed and published the data of 7340 information operations accounts that are employed for astroturfing campaigns and were centrally managed in a coordinated manner~\cite{grossman2020political}. We inspected this data and identified 77 accounts that were tweeting lexicon tweets to push fake trends. Second, we set up a honeypot account and expose its credentials to a phishing website that was advertised by the fake trends. Our account became part of the botnet and participated in 563 attacks by automatically posting and deleting lexicon tweets in 6 months. Please refer to our previous study for more details~\cite{elmas2021ephemeral}. 

\Figref{fig:elonmusk} shows an example of a fake trend "ElonMusk TvitterıMafetti" (ElonMusk RuinedTwitter)\footnote{https://getdaytrends.com/trend/ElonMusk\%20TvitterıMafetti} created by tweets observing these two patterns. 200 bots mentioned the trend in lexicon tweets in a coordinated manner within the same minute. They later deleted them and stayed silent. Despite being an obvious spam campaign, the trend, made it to the top spot in Turkey and ranked 8th globally.

\section{Data Collection Methodology}
\label{sec:data_collection}

As the attack tweets employed in ephemeral astroturfing are deleted in a short time period, it is not possible to collect them using keyword-based queries that are performed retrospectively. This is because by the target keywords listed as trends and are known to the public, the bots have already deleted their tweets. Thus, we employ two methods that collect the data in real-time. 

First, we employ the 1\% real-time sample provided by Twitter. We collect this data from the Internet Archive which stores this data between 2011 and 2022~\cite{archive}. This dataset contains the deleted tweets and the deletion notices which announce when a tweet is deleted. As our focus is the fake trends by bots, we only kept the tweets that contain trending keywords and discarded the rest. To do this, we employed our collection of Twitter trends in Turkey which contains all trends since 2013. Finally, we only used the deleted tweets during the process as deletions are the key parts of the attack. 

\begin{figure}
    \centering
    \includegraphics[width=\columnwidth]{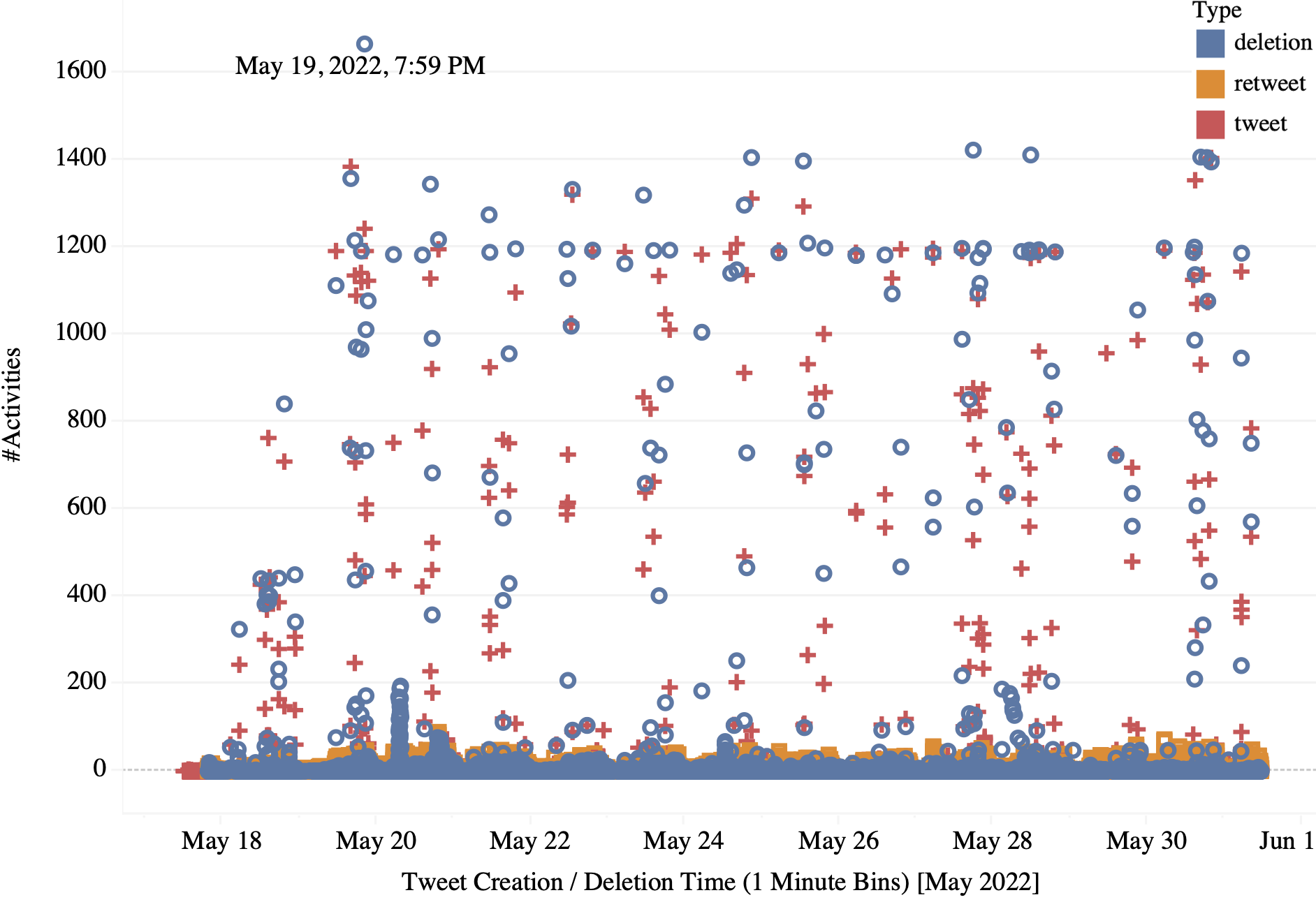}
    \caption{Two weeks of activity by the bots we tracked in real-time. The volume of activity is aggregated per minute. Blue circles denote the deletions, red crosses denote the tweets and orange squares denote the retweets. We observe that the bots are coordinated in their tweets and deletions. For instance, on May 19th, 2022, 1666 bots mass deleted their tweets at 7:59 PM, within the same minute.}
    \label{fig:sample_bottracker}
\end{figure}

As the 1\% sample may yield a low recall of tweets used in the attacks, we complement this data with a second strategy. We tracked a sample of 5000 bots that we observed to be promoting fake trends by posting lexicon tweets and later deleting them in real time using Streaming API. That is, we collected an initial sample of 753 accounts from the 1\% random sample between August 2021 and January 2022. We then expanded this dataset by tracking those accounts and collecting the data of the trends this initial set promote, and then manually identifying the other astrobots promoting those fake trends. \Figref{fig:sample_bottracker} shows a 2 weeks sample of the tracked bots' activity. Using this activity data, we detected the fake trends they promote at the moment they started to promote them using the methodology outlined in~\ref{sec:detectfakes}. We then collected the other tweets by bots using Search API using the fake trend as the input. As we do this before the accounts delete their tweets, we could capture full activity promoting which enriches our bot dataset. 

\section{Detection Methodology}

Our main objective in this work is to detect the astrobots that are employed in the attacks. Thus, we first detect the fake trends created by those attacks. Then we detect the attack tweets that are employed in the attacks. We finally detect the bots that author the attack tweets. We also describe the ground truth used for each task. 

\subsection{Fake Trend Detection}
\label{sec:detectfakes}

We replicate the methodology we provided earlier~\cite{elmas2021ephemeral} to detect fake trends using the 1\% sample as the data. 

\noindent\textbf{Ground Truth:} We use the ground truth we proposed earlier. That is, we annotated a trend as fake, created by an ephemeral astroturfing attack if they originated (i.e., first mentioned in) attack tweets, which are deleted lexicon tweets in our case. We took a random sample of 5 trends per day between 2017 and 2019, 4255 trends in total. We annotated the trends by inspecting the first 10 tweets mentioning them. We annotated 843 trends as fake trends. We split the training and the test set with an 80\%:20\% ratio. The test data starts from trends in February 2019. The training set contains 648 positives and 2,756 negatives while the test set contains 195 positives and 656 negatives.

\noindent\textbf{Detection:} We used our previous fake trend detection model which relies on a Decision Tree with a depth of two, i.e., two rules. First, the trend should be mentioned by at least 4 tweets that are detected to be attack tweets. Since this is based on a 1\% sample, this is roughly equal to 400 attack tweets in the complete data. Secondly, more than 45\% of the detected attack tweets must be deleted. Although this detection is simple, it is very effective because the behavior is very unique and it is easy to distinguish it from organic trends. The method reports a 99.7\% 5-fold cross-validation score, 100\% precision, 98.9\% recall, and 99.4\% F-score on the test data~\cite{elmas2021ephemeral}. We detect the attack tweets using a rule-based classifier to check if a tweet is a lexicon tweet or not (See \Secref{sec:attack_tweet_detection}). 

\subsection{Attack Tweet Detection}
\label{sec:attack_tweet_detection}

We classify if a single tweet is an attack tweet, i.e., part of an attack or not. We have two different data sources with different granularity. Thus, we propose a ground truth for each data source. 

\noindent\textbf{1\% Sample:} 1\% random sample of all tweets that is assumed to be unbiased. We use it to perform the longitudinal analysis. 

We used the ground truth we proposed earlier~\cite{elmas2021ephemeral} which consists of the first 10 tweets promoting a random sample of 127 fake trends between January 2017 and September 2019, 16742 tweets in total. 8446 of those tweets were attack tweets (i.e., a series of lexicon tweets that are later deleted) while the 8296 non-were non-attack tweets (i.e., other tweets mentioning the fake trend but were not deleted lexicon tweets). We use 80\% of the data (corresponding to 101 trends) as the validation set and the rest as the test set.

This sample suffers from low recall on attack tweets as it is limited to 1\%, e.g., the adversaries employ 400 attack tweets and only 4 show up in the sample. Thus we complement it with an additional data source to increase the recall of attack tweets and the detected bots. 

\noindent\textbf{Full:} Complete data of tweets mentioning trends that we detected in real-time by tracking the 5000 bots. We collected this data in real time using the Search API before the data is deleted. The data yields a higher recall on attack tweets but does not guarantee a full recall on fake trends as it is limited to 5000 bots. 

Since this data has different granularity than the previous one (100\% vs 1\%), the attack tweet detection models' performance will differ. Thus, we propose an additional ground truth dataset to test our detection models. We annotated the data of 72 trends which consists of 37,728 tweets between 17 May and 31 May 2022. We used the same annotation approach: we annotated a series of uninterrupted lexicon tweets that promote a trend that is later deleted as attack tweets. We annotated 11,091 tweets as attack tweets and 26,637 tweets as non-attacks in total. We use the data from 60 trends (9,226 attack tweets and 21,047 negatives) as the validation set and the data from 12 trends (1,865 attack tweets and 5,590 negatives) as the test set. 

We experiment with two unsupervised classifiers. We validated and evaluated them on the two ground truth datasets. 

\noindent\textbf{Lexicon Classifier:}: We classify attack tweets by annotating if they appear to be lexicon tweets or not. In our previous work, we curated the following rules based on the tweets posted from our honeypot account:

\begin{enumerate}
\item Has between 2-9 tokens, excluding emojis.

\item Begins with a lowercase letter. (False negatives were proper nouns from the lexicon.)

\item Alphabetic characters only, except for parentheses or emojis. 
\end{enumerate}

\noindent\textbf{Isolation Forest:}: We observe that the attacks are anomalous signals that consist of a single spike in time as seen in ~\Figref{fig:elonmusk}. Thus, we employ content-agnostic anomaly detection and classify each tweet that is part of an anomaly as an attack tweet. We use Isolation Forest~\cite{liu2008isolation} as our early experiments showed that it is faster and performs better than its competitor, Local Outlier Factor~\cite{breunig2000lof}. Precisely, we group the tweets per minute windows and compute the anomalous windows where there is a spike in the volume of tweets per minute using this algorithm. The algorithm takes one parameter, the outlier factor, which we empirically tune using the validation set and report the final result on the test set. 

The experimental results are summarized in Table \ref{tab:clf_results}. The Isolation Forest suffers from recall when we use the 1\% sample as the data source. This is because this sample is insufficient to create anomalies for the same attacks. On the other hand, it performs better in the complete data as the anomalies are clear as in \Figref{fig:elonmusk}. Thus, we use the lexicon classifier to annotate and analyze the accounts in the 1\% sample and Isolation Forest on the latter dataset. 

\begin{table}[]
\caption{The classification results of individual attack tweets}
\begin{tabular}{|l|c|c|c|c|}
\hline
Metric         & \multicolumn{1}{l|}{Lex. (1\%)} & \multicolumn{1}{l|}{Lex. (Full)} & \multicolumn{1}{l|}{Iso F. (1\%)} & \multicolumn{1}{l|}{Iso F. (Full)} \\ \hline
Prec.-Val  & 0.99                               & 0.975                               & 0.979                             & 0.970                              \\ \hline
Recall-Val     & 0.937                              & 0.733                               & 0.661                             & 0.977                              \\ \hline
F1-Val         & 0.967                              & 0.837                               & 0.789                             & 0.974                              \\ \hline
Prec.-Test & 1.0                                & 1.0                                 & 0.956                             & 0.978                              \\ \hline
Recall-Test    & 0.808                              & 0.482                               & 0.699                             & 0.981                              \\ \hline
F1-Test        & 0.893                              & 0.652                               & 0.807                             & 0.979                              \\ \hline
\end{tabular}
\label{tab:clf_results}
\end{table}

\subsection{Attack Tweet Detection Results}

We first run the lexicon classifier on the 1\% sample to mass-annotate lexicon tweets. We found 1,171,487 deleted lexicon tweets from 265,427. We then run the fake trend detection method and found 29,271 fake trends and 28,359 negatives. We then deleted the data of negatives (129k tweets). The final data is \textbf{1,042,551} deleted lexicon tweets from \textbf{195,116 bots.}

We found that 43,724 of those accounts are already suspended, and 36,733 accounts are "Not Found" (maybe deactivated by the adversary). 114,659 of the accounts, which make up 59\% of the dataset, still exist on the platform. The rest of the analysis is based on the activity patterns of those bots. 

We also run Isolation Forest on the data of 1,456 trends from 2022 for which we could collect their data using Search API. This process revealed 37,195 accounts and increased the dataset by 17,800 accounts. The final data we share consists of \textbf{212,916 bots.}

\section{The Prevalence of Fake Trends}
\label{sec:faketrendprevalence}

\begin{figure}
    \centering
    \includegraphics[width=\columnwidth]{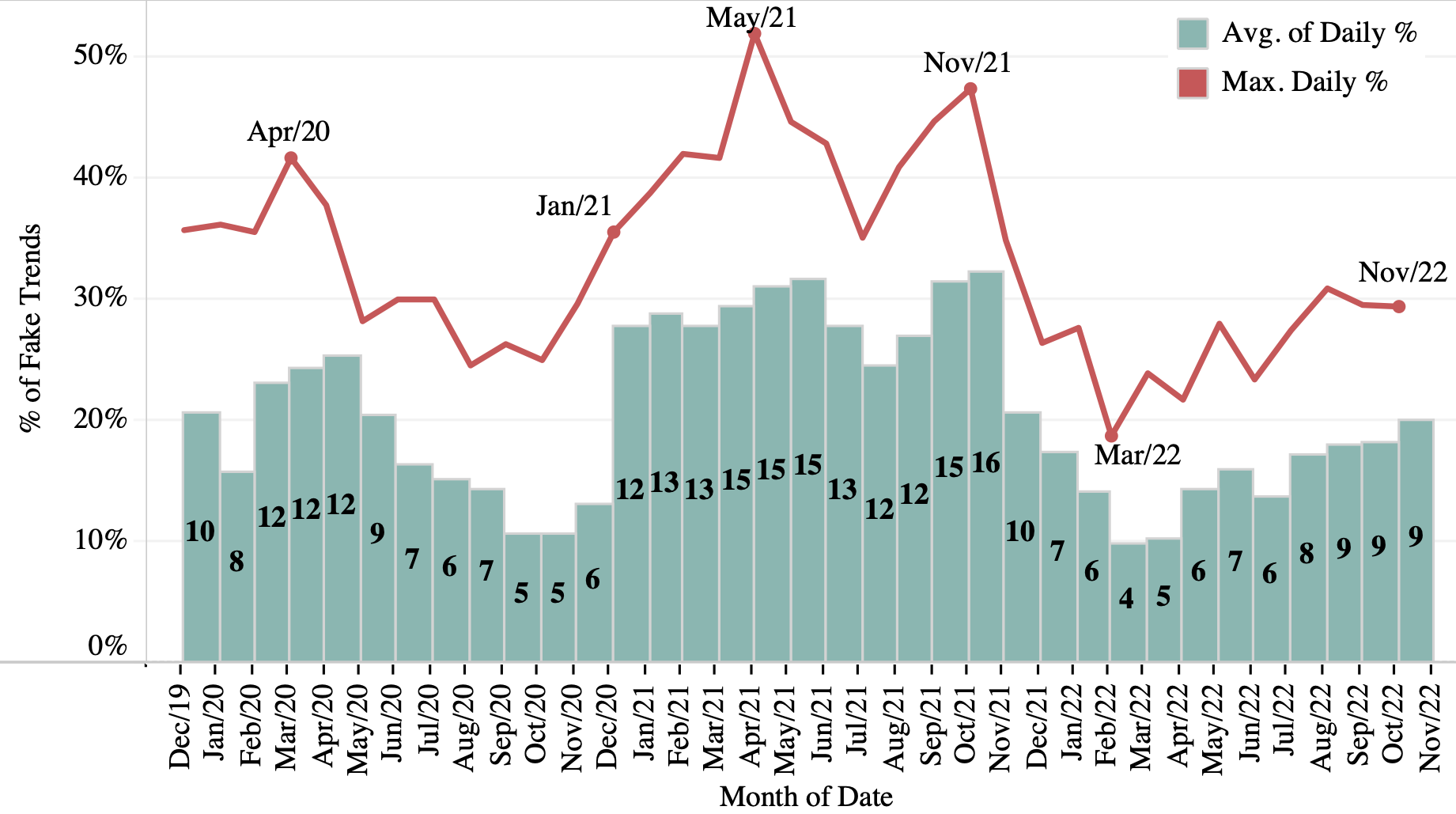}
    \caption{The share of fake trends in the top 5 trends each day. The bars indicate the average of the percentages of fake trends each day, while the numbers on the bar indicate the average number of fake trends each day within the month. The lines indicate the maximum percentage of fake trends observed in a day within that month. We observe two periods of decline in fake trends, beginning in May 2020 and November 2021, followed by a steady rise after 6 months.}
    \label{fig:fake_trend_prev}
\end{figure}

We previously reported that the average daily percentage of fake trends within the top 5 trends was 47\% over three months between June and September 2019. We observe that this value decreased to 21\% in 2020, which we suspected to be due to Twitter's intervention around August 2019. Furthermore, the fake trends had another decline, from 25\% in May 2020 to 11\% in November 2020. We did not observe a high volume of new bot activity or suspension between August 2019 and November 2020. Thus, this decline may be due to Twitter proactively preventing bot activity. Or, it may be due to low demand for fake trends as it coincides with the lockdown. We observe another increase that started in January 2021, which is accompanied by a new bot-net being active at that date. Between January 2021 and November 2021, the average percentage of fake trends in a day was 30\%. With Twitter purging the bots around November 2021, the share of fake trends declined, dropping to 10\% in March 2022. However, it has been steadily increasing since and has risen to 20\% in November 2022.

By April 7, 2023, prior to a month before the 2023 Turkish election, the attacks continue. The percentage of fake trends has been oscillating between 35\% to 9\% (weekends). Moreover, we have observed they have been used to push more impactful slogans that represent political manipulation campaigns. 

Our caveat here is that this analysis is on 1\% of data; there may be many trends for which the Twitter API did not provide any associated tweet within this sample. This prevents us from classifying such trends as fake and underestimates the share of fake trends. However, the tendency of the fake trends to increase and decrease should not be affected by the volume of the data, unless Twitter introduced changes to API.

\section{Account Activity, Creation and Suspension Patterns}
\label{sec:astrobotcreationsuspension}

\begin{figure*}[!htb]

    \begin{subfigure}{.33\textwidth}
        \centering
        \includegraphics[width=\linewidth]{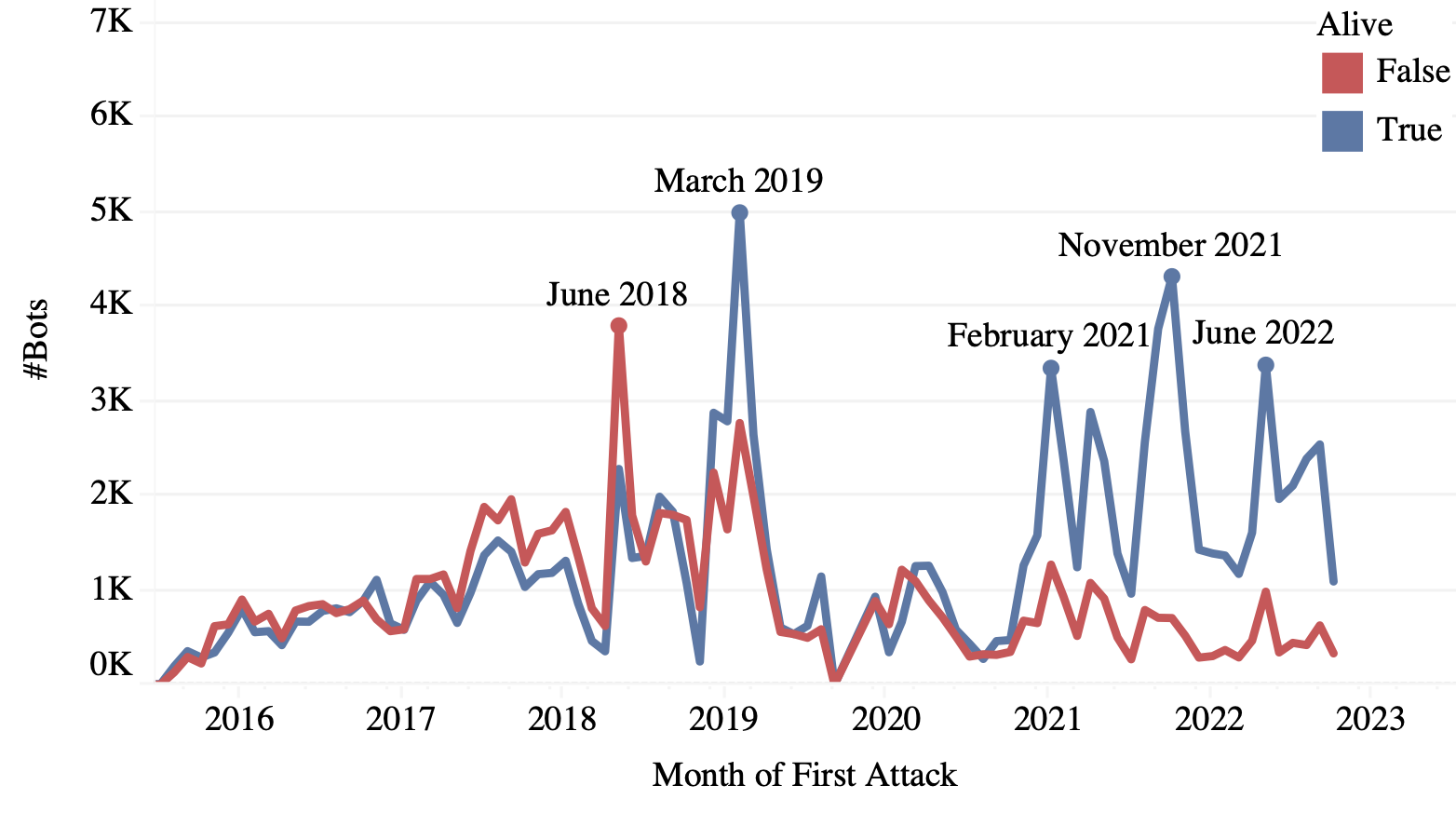}
    \end{subfigure}
    \begin{subfigure}{.32\textwidth}
        \centering
        \includegraphics[width=\linewidth]{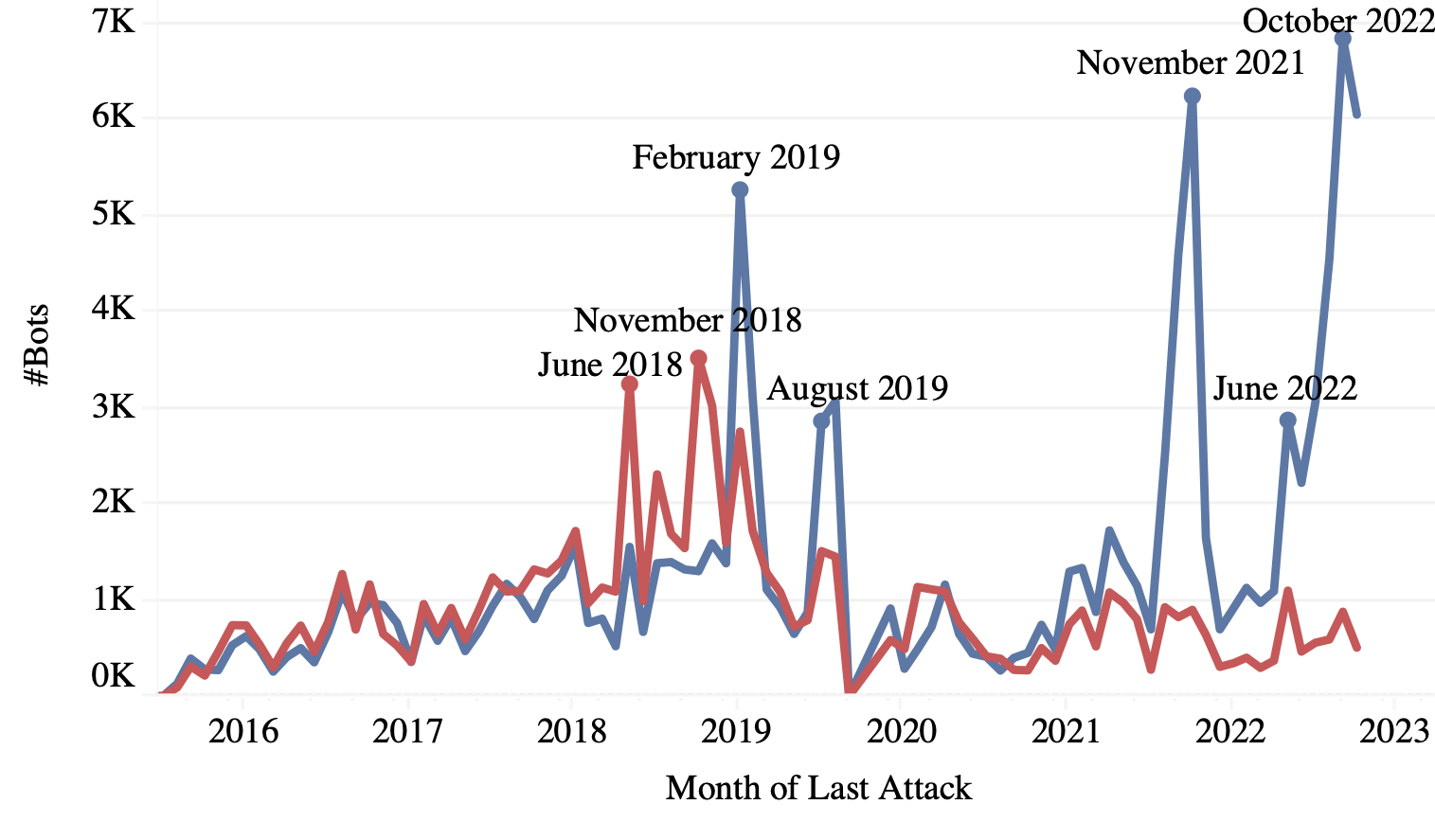}
    \end{subfigure}
    \begin{subfigure}{.33\textwidth}
        \centering
        \includegraphics[width=\linewidth]{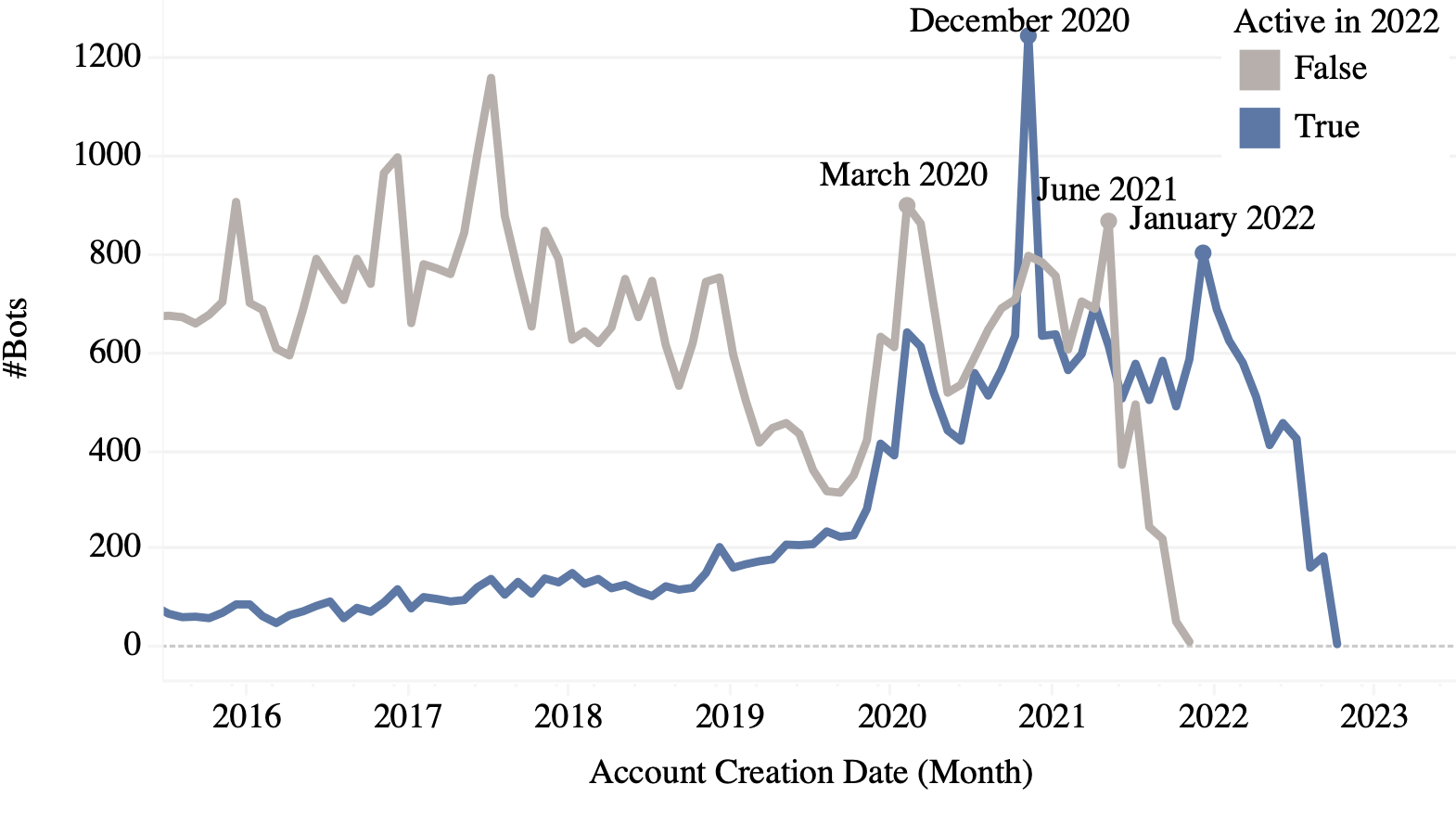}
    \end{subfigure}
    \caption{The number of bots per the month they are first seen attacking (left), last seen attacking (middle), and their creation date (right), if they still exist on the platform as of February 2023. We observe that Twitter often purges accounts in bulk by removing them or making them cease to operate. As a response, the adversaries start to operate new accounts simultaneously. The accounts are also bulk created, but they do not start to operate immediately.\label{fig:longitudinal bots}}

\end{figure*}

Our objective is to understand Twitter content moderation practices which may be the factor in the decline of the attacks. We also aim to analyze how the adversaries react to Twitter removing their users. Thus, we collected the user's current profile in February 2023 to understand whether the account is suspended, deactivated, or still alive. We use three types of signals to capture the activity and suspension patterns of bots. We first use creation dates, as a high volume of bots created within the same date indicates that they are bulk-created by an adversary and may be used as part of the same bot-net. We then use the first and the last date the bot is employed for an attack. Past work revealed that the accounts that were employed by a black market started to operate simultaneously although the accounts were not bulk-created and may have become part of the bot-net due to getting compromised or purchased. They also ceased activity at the same time even though they were not suspended which indicates that Twitter imposed a soft ban on the accounts in that bot-net~\cite{elmas2022characterizing}. As the accounts we investigate are also likely parts of fake trend black markets, the first and last dates they attack capture the time they started and ceased operations. \Figref{fig:longitudinal bots} shows the number of bots versus the month of the first and last seen attacking, and their creation dates (if they are not suspended). We now interpret the results. 

\subsection{Temporal Patterns in Attacks' Start}
We observe that over 4000 bots started to operate on five key dates: June 2018, March 2019, February 2021, November 2021, and June 2022. June 2018 is the month when the 2018 Turkish General Election took place, while March 2019 is the month the 2019 Turkish Local Election took place. The latter three dates do not observe any big political events. We manually inspected the fake trends in these three months but did not observe any recurring fake trends on a political or another significant event either. The surge of new bots starting to operate around these dates may be rather a response to Twitter purging bots in bulk; the adversaries may have created or converted new bots and started to operate them simultaneously. We provide evidence for this hypothesis in the next subsection.

\subsection{Temporal Patterns in Attacks' End}

\begin{figure}
    \centering
    \includegraphics[width=0.73\columnwidth]{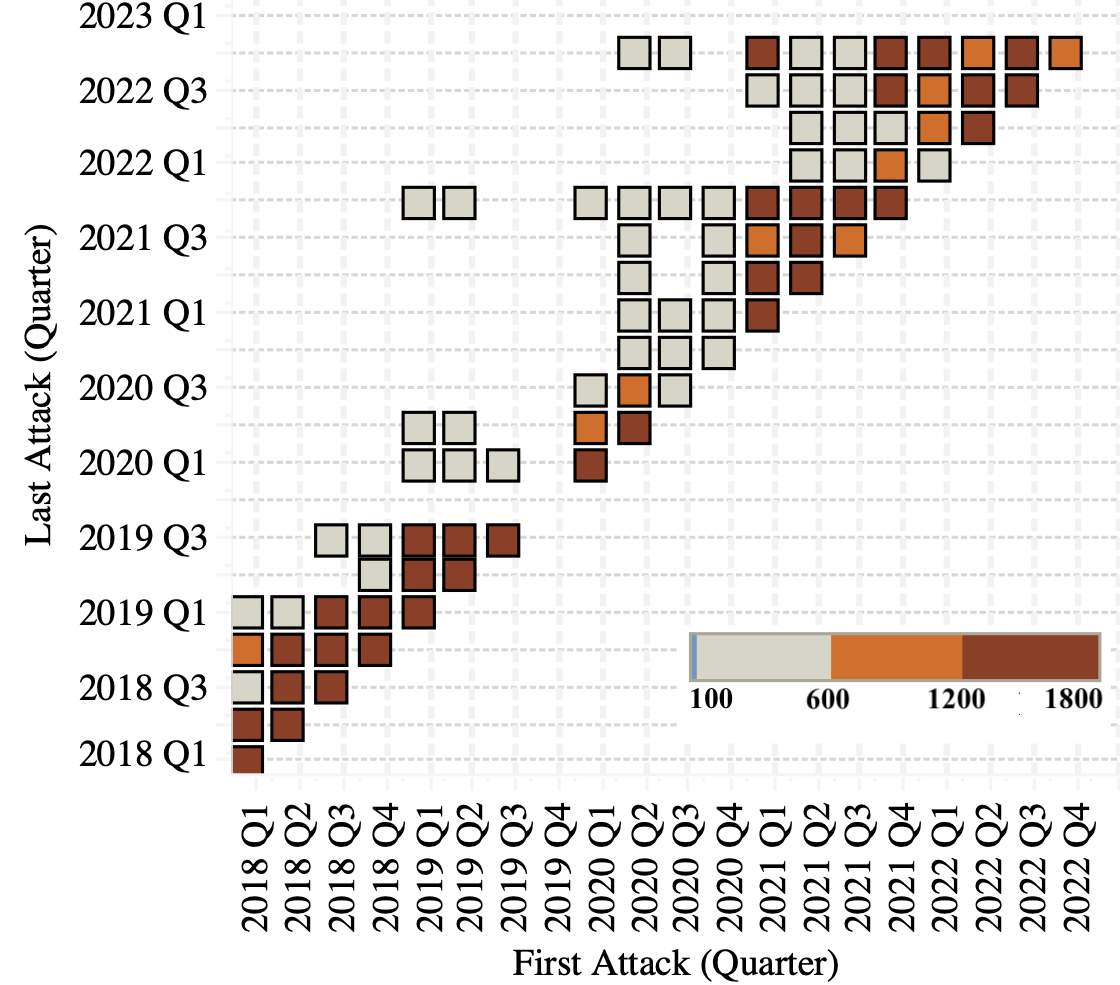}
    \caption{The number of bots attacking for the first time (x-axis) versus last time (y-axis). Each square on the graph represents at least 100 bot-nets that started and ceased operations within the respective quarter of the year.}
    \label{fig:first_last}
\end{figure}

Multiple communities of bots ceased activity around June 2018, November 2018, February 2019, August 2019, and November 2021. There is also a new wave of bots ceasing activity starting in June 2022 which coincides with Elon Musk's purchase of Twitter. In the first three periods, the accounts are mostly removed while in the latter three, they mostly ceased activity without removal. We inspected each case to check when the accounts were created or started to operate to describe the arms race between the adversaries and Twitter. Figure~\ref{fig:first_last} provides a summary of the bot-nets' time of activation and cessation.

85\% of the accounts that ceased activity in June 2018 started to operate \emph{before} June 2018, meaning that they were able to manipulate the elections before Twitter removed them. This may also be the reason why there was a new wave of accounts starting to operate in June 2018, which were suspended later. If this is the case, the adversaries could replace the bots in no time, showing that only removing bots is not an effective method. 95\% of the accounts that started to operate in this new wave was removed between July 2018 and February 2019 (hence the peak around November 2018), which means Twitter acted rather quickly against this bot-net.

Almost all (91\%) of the accounts that ceased to operate in February 2019 started to operate after June 2018. However, the majority (62\%) started to operate in January and February 2019. This means that Twitter primarily targeted the new wave that started to operate after the purge in November 2019, which was a rather quick intervention compared to others.

As a response to the purge in February 2019, new accounts started to operate in February and March 2019. The purge in August 2019 targeted mainly this new-bot-net (75\%).

A new wave of bots started to operate around February 2021 and afterward but they ceased activity in November 2021. The purge in October and November 2021 was huge: more than 10,000 accounts ceased operation. A significant portion of those accounts (42\%) started to operate within those two months, and (46\%) started in and after January 2021 until October 2021. However, this purge was still not final; 22\% of the accounts that ceased operation in October 2022 started to operate before October 2021, meaning that they survived the purge in October and November 2021.

This analysis shows us that Twitter  typically removes accounts in large numbers, by either suspending them or making them cease to operate. Consequently, adversaries replace those accounts with new ones, resulting in new batches of bot accounts starting to operate simultaneously. Our hypothesis is that they do this not by creating new accounts, but by keeping a list of backup accounts or buying accounts from a provider that has already created accounts in bulk to replace the purged ones. We provide evidence for this by analyzing the creation dates of the accounts in the next subsection.

\subsection{Temporal Patterns in Account Creation}
\label{sec:account_creation}

\begin{figure}
    \centering
    \includegraphics[width=0.9\columnwidth]{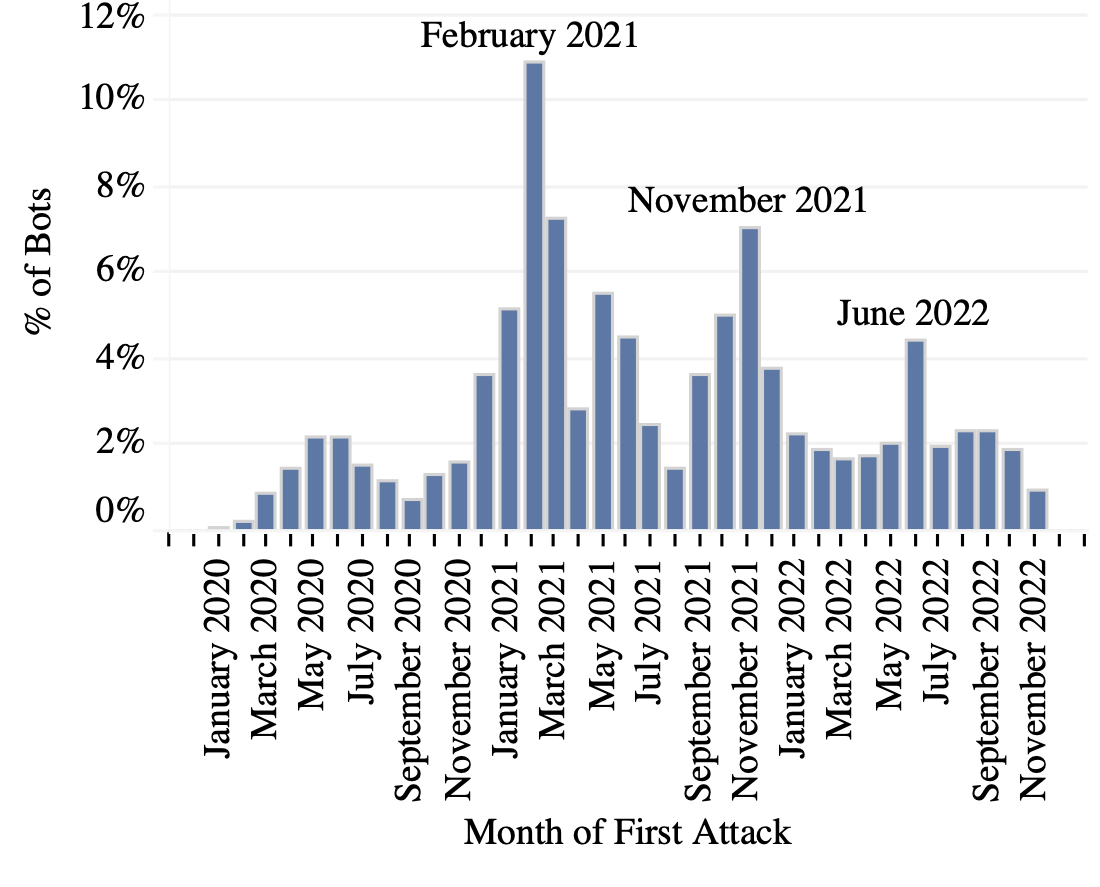}
    \caption{The accounts that are created in 2020 and their month of the first activity. Despite being created in 2020, they mainly started to operate in February 2021, November 2021, or June 2022 as a response to the Twitter purge(s).}
    \label{fig:accounts_created_in_2020}
\end{figure}

\begin{figure}
    \centering
    \includegraphics[width=0.8\columnwidth]{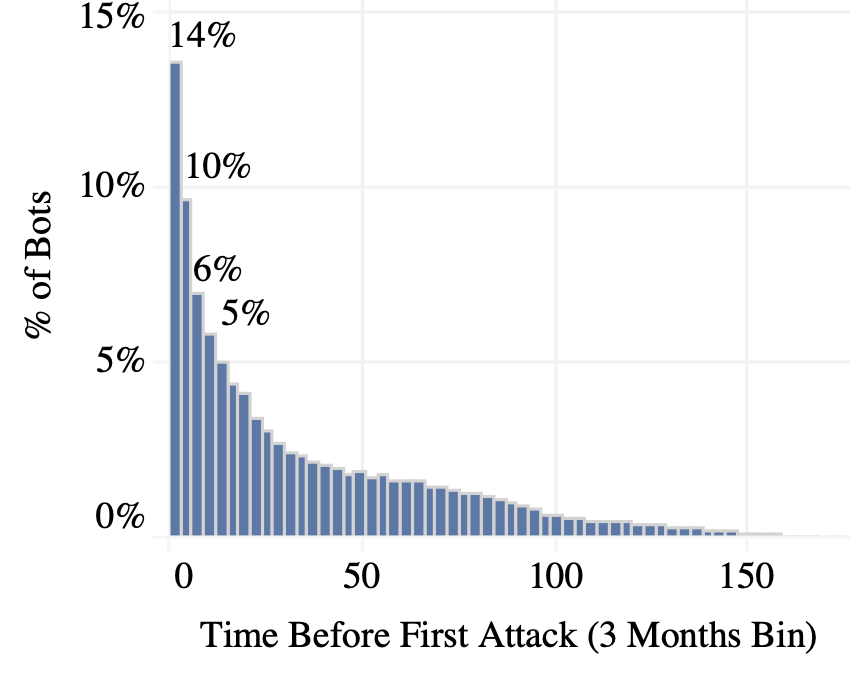}
    \caption{The time between the account creation and the date of the first attack, binned by 3 months windows. Many accounts start to participate in attacks after a year. }
    \label{fig:time_before_first_attack}
\end{figure}

To analyze when the accounts are created, we computed the number of accounts created within the same month, as shown in \Figref{fig:longitudinal bots}. We observe that the account creation dates do not necessarily reflect the date the bots start to operate. The accounts that were not removed as of February 2023 were mostly created between January 2020 and February 2022. The account creation dates spiked around March 2020, December 2020, June 2021, and January 2022. The majority of the accounts that are created after January 2020 (57\%) were still active in 2022, indicating that they have survived the purge in November 2021. This is because most of those accounts started to operate \emph{after} the November 2021 purge. We plot the month of the first attack of the accounts created in 2020 in \Figref{fig:accounts_created_in_2020} to show this. We observe that the majority (84\%) of the accounts started to operate in and after 2021, with their start of operations peaking around Twitter purges in February 2021, November 2021, and June 2022 purges. This supports our hypothesis that the adversaries create and maintain a database of bots and use them as sleeping agents, and replace them when Twitter purge the bots they currently operate. Alternatively, they purchase more accounts when Twitter removes the accounts they currently employ from a provider, who bulk-created accounts. To further verify that this is the case, we compute the time before the first attack for all accounts in \Figref{fig:time_before_first_attack}. Among the accounts that are still not suspended, the majority of the accounts (65\%) started to operate after a year of account creation. Only 14\% started to attack within three months. Adversaries may be using accounts after a year as a strategy against Twitter's content moderation and soft-censor practices. Twitter may be filtering new accounts with their quality filter~\cite{qualityfilter}, causing adversaries to rely less on new accounts.

\section{Bot Characteristics}
\label{sec:botcharacteristics}

We describe three characteristics of bots in this dataset which may provide insights into future bot detection research. 

\subsection{Number of Attacks}

\begin{figure}
    \centering
    \includegraphics[width=0.65\columnwidth]{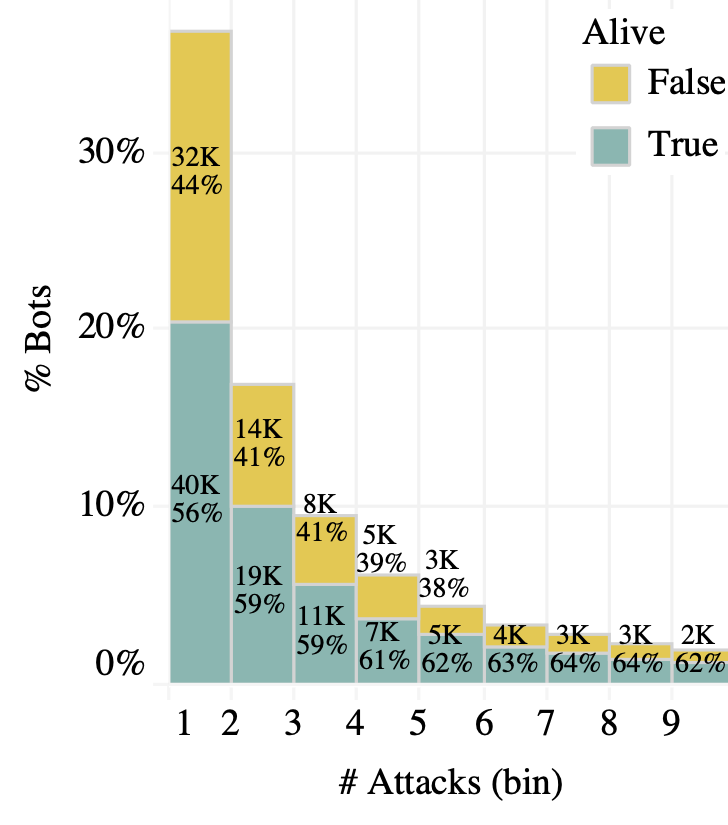}
    \caption{The number of attacks per account. Cyan indicates that the accounts still exist on the platform by February 2023, which is roughly 60\% regardless of the number of attacks in the account involved.}
    \label{fig:total_attacks}
\end{figure}

We characterize the number of attacks they participated in. We do this to understand how many attacks the adversaries can use a single account and how this affects getting suspended. As this is based on the 1\% data, the numbers are underestimated, but the patterns will likely remain the same. \Figref{fig:total_attacks} shows the histogram of accounts binned by their number of attacks up to 10. As expected, the majority of users (65\%) attacked less than 4 times and only a minority (15\%) attacked at least 10, with the maximum being 193.  

However, the important finding of this analysis is that there is no correlation between the percentage of being removed and the number of attacks. The suspension probability is uniformly distributed across the accounts binned by their number of total attacks. Roughly 40\% of the accounts are suspended or deactivated whether they were detected to be attacked once, twice, or more than 20 times. This may be another evidence that Twitter does not take the bots' hyper-activity into account; they are more likely to purge the bots in bulk at certain times, as we described in the previous section. 

\subsection{Silent Attacks}

A crucial aspect of this dataset that should not be overlooked is that many bots in this dataset are silent; they either do not have any tweets on their profile or the account stopped tweeting long ago. However, the accounts are silently used to create fake trends; they post tweets promoting a fake trend and immediately delete them. This makes their accounts appear to be inactive and makes it hard for bot detection methods to detect them.

\begin{figure}
    \centering
    \includegraphics[width=0.9\columnwidth]{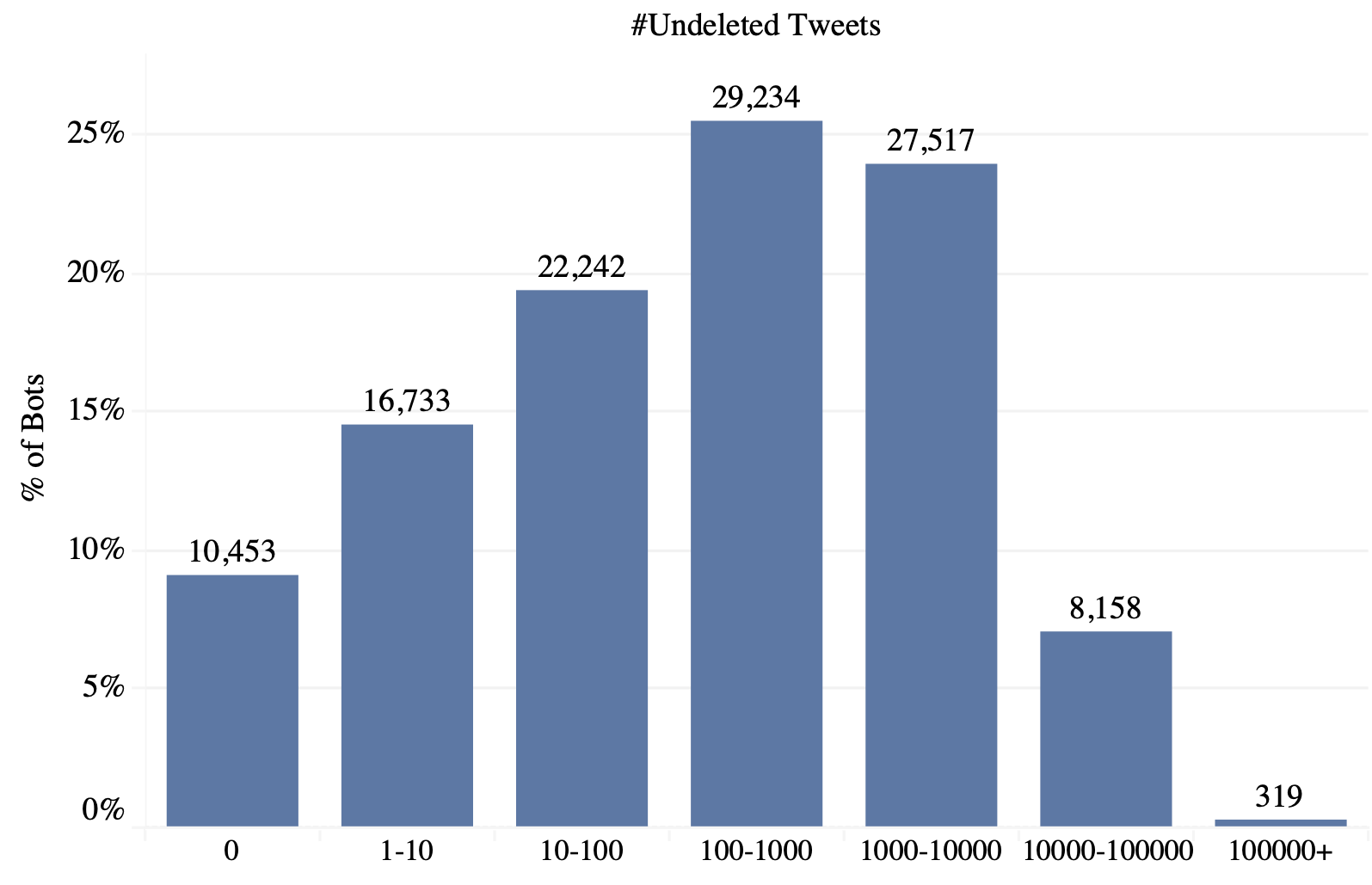}
    \caption{The number of accounts per their undeleted tweets.}
    \label{fig:undeleted}
\end{figure}

To present evidence for the former, we show the total number of undeleted tweets (including retweets) per account in \Figref{fig:undeleted}. We observe that there are 10,000 accounts (9\%) with no tweets remaining on the profile, and the majority of the profiles have less than (67\%) 1000 tweets in their profile.

\begin{figure}
    \centering
    \includegraphics[width=0.8\columnwidth]{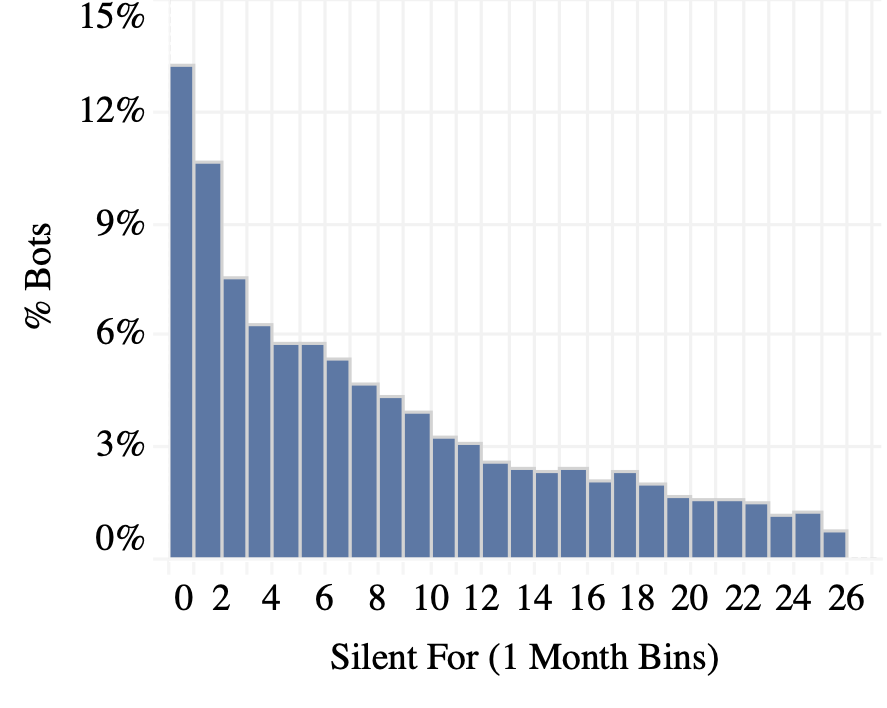}
    \caption{The number of months between the last attack and the last undeleted tweets. Many accounts continued to attack despite they appear to stop tweeting a long ago.}
    \label{fig:silentforhist}
\end{figure}

To show the latter case where the account is still actively used as a trend bot although the account stopped tweeting, we compute the number of days between the date of the last attack and the date of the last tweet still present on the profile. \Figref{fig:silentforhist} shows the result. In fact, 87\% of the bot accounts are silent for at least 1 month; they promoted a fake trend and deleted tweets despite the account appearing to be not in use or active. 

This has significant implications for bot research. First; the detection methods should not only rely on tweeting activity as accounts that do not tweet may be actively used as part of a bot-net. Second; the bot researchers should rely on real-time data collection to be able to collect such ephemeral activity. A retrospective collection of social media data would not be able to account for the activity of bots in this dataset. Third; there are many other bot functions that do not involve tweeting such as engaging in fake followings and likes. We leave investigating if those bots observe such behaviors in future work.

\subsection{Usage of Default Profile Handles}
In \Secref{sec:account_creation} we revealed that many accounts are mass-created recently to replace the bots that got suspended or locked. We further show that this is the case by analyzing the usage of default profile handles. Twitter assigns default profile handles to users which can be changed manually. Those handles sometimes contain an anomalous number of numerics, e.g., @realdonald12345678. We found that many, if not the majority of, users observe this phenomenon. Precisely, roughly 7.5\% of the bots (8500 accounts) were subject to such lazy account creation and continue to use their handle as such which consists of 8 numbers in their handles. This may suggest that such behavior can indicate inauthentic accounts that are created to later use as bots.

\section{Dataset}

\label{sec:dataset}

Our dataset includes full tweeting/deletion activity found in the 1\% sample between 2015 (when the attacks started) and 2022. We additionally include the complete activity associated with 1,456 trends, collected using Search API in 2022. We also include the full activity of 5000 users we tracked between May 2022 and November 2022. The dataset can verify or improve bot detection based on profile features or other behaviors. Furthermore, as these bots coordinate when they attack Twitter trends to promote a fake trend, they can be used as an auxiliary dataset to verify and further improve coordination-based detection methods. 

We are releasing this dataset ahead of the 2023 Turkish elections. Thus, our dataset may enhance the analysis of polluters of public discourse on the elections. 
\section{Conclusion}

In this work, we detected, analyzed, and propose a dataset of bots targeting Turkish trends with unique and anomalous behavior. We briefly discuss the caveats of this work and its generalizability.

\subsection{Caveats} Like any other dataset for Twitter bots, this dataset should be used with care. As our analysis shows, the accounts may cease operation and exit the botnet. We provide their activity and profiles during the time they are part of the botnet. However, we recommend researchers track the accounts and confirm that they are still part of the botnet (i.e., actively post and delete tweets that mention trends) if they opt for collecting and using the account's up-to-date profiles. Additionally, Twitter bots often have a primary function and may not observe other kinds of bot behaviors. For instance, fake followers may not engage in retweeting activities~\cite{cresci2015fame} or retweet bots may not be overactive and post tweets in high volumes despite maintaining an anomalous level of retweeting activities~\cite{elmas2022characterizing}. Similarly, some astrobots may not be involved in other malicious behavior even though they fit the definition of a bot, i.e., they are controlled by software at least in part. We recommend researchers to analyze if they observe behavior other than the trend manipulation (e.g., fake following) if that behavior is the one they aim to detect.

\subsection{Generalizability} Although we mass-annotated the fake trends and bots by detecting a unique and anomalous paper, the rest of the analysis on activity and suspension patterns can be applied to any bot-net, and can inspire future work on the analysis of bot-nets. 

\section{Acknowledgements} This work was supported in part by DARPA (grant HR001121C0169).

\bibliographystyle{ACM-Reference-Format}
\bibliography{sample-base}

\appendix

\end{document}